\documentclass[]{aa} \usepackage{graphics,amsmath,placeins,threeparttable}

\begin{document}

\thesaurus{06(02.01.2; 08.02.1; 08.09.2; 08.14.1; 13.25.5)}

\title{On   the    X-ray   fast-time   variability    of   \object{Sco~X-2}
(\object{GX~349+2})}

\author{Paul~M.~O'Neill\inst{1}\and              Erik~Kuulkers\inst{2,3}\and
Ravi~K.~Sood\inst{1} \and Tadayasu~Dotani\inst{4}}

\offprints{P.~M.~O'Neill}

\institute{School~of~Physics,  Australian~Defence~Force~Academy, University
           of    New     South    Wales,    Canberra~ACT~2600,    Australia
           (p.m.oneill@adfa.edu.au~~r.sood@adfa.edu.au)                 \and
           Space~Research~Organization~Netherlands, Sorbonnelaan 2, 3584 CA
           Utrecht,           The~Netherlands~(erikk@sron.nl)          \and
           Astronomical~Institute, Utrecht~University, P.O. Box 80000, 3507
           TA            Utrecht,            The~Netherlands           \and
           Institute~of~Space~and~Astronautical~Science,
           1-1~Yoshinodai~3-chome,   Sagamihara-shi,   Kanagawa   229-8510,
           Japan~(dotani@astro.isas.ac.jp)}

\date{Received / Accepted }
        
\titlerunning{On  the X-ray  fast-time variability  of  Sco~X-2 (GX~349+2)}
\maketitle

\begin{abstract}

We  have analysed  archived Ginga  data  on the  Z source  \object{Sco~X-2}
(\object{GX~349+2}).  We  present the  first detailed investigation  of its
X-ray  fast-time variability, as  a function  of position  in the  Z track.
During the two-day observation over  the period 5--7 March 1989, the source
was in  the so-called flaring branch,  and the lower part  of the so-called
normal branch.  We  found broad peaked noise with  a centroid frequency and
width of  $\sim$4--7~Hz and $\sim$6--12~Hz respectively.   The peaked noise
was strongest  in the lower flaring  branch, with a  maximum fractional rms
amplitude of $\sim$3~$\%$.   We conclude that it is  not a manifestation of
atoll source  high frequency noise, as  had been suggested,  and compare it
with the power spectral features seen in other Z sources.  We find that the
peaked noise is markedly different to the quasi-periodic oscillations found
in the  normal and flaring  branches of \object{Sco~X-1}; however  it bears
some resemblance to that seen  in the flaring branch of \object{Cyg~X-2} at
low overall intensities.


\end{abstract}

\section{Introduction \label{int}}

\object{Sco~X-2}  (\object{GX~349+2})  is  a  neutron star  low-mass  X-ray
binary and Z source.  Z sources trace out a characteristic `Z' shaped track
in an X-ray colour-colour  diagram (CD) or hardness-intensity diagram (HID)
consisting  of  a  `horizontal'  branch  (HB),  `normal'  branch  (NB)  and
`flaring' branch  (FB) (Hasinger~\&~van~der~Klis~\cite{hv89}; hereafter HK;
van~der~Klis~\cite{v95}).  The  orientation of each  branch in a CD  or HID
depends upon  the choice of energy bands.   Mass-accretion rate ($\dot{M}$)
is thought to increase  from the HB, down the NB and  up the FB.  The other
traditional    Z   sources    are:    \object{Sco~X-1},   \object{GX~17+2},
\object{GX~340+0}, \object{Cyg~X-2}, and \object{GX~5-1}.

A class of  objects that are related to Z sources  are the `atoll' sources.
Atoll  sources   trace  out  a  curved   track  in  a   CD,  with  inferred
mass-accretion rate increasing along  the track.  Atoll sources are thought
to have lower magnetic field strengths, and lower mass-accretion rates than
Z sources (HK).

The X-ray fast timing properties of  Z and atoll sources are different, yet
related.  The  features observed in the  power spectra of  both classes are
normally correlated  with the  position of the  source in  a CD or  HID.  Z
sources exhibit very low frequency noise (VLFN), low frequency noise (LFN),
high frequency  noise (HFN), and quasi-periodic  oscillations (QPOs).  QPOs
with frequencies in the range 15--60~Hz (called HBOs) and 200--1200~Hz (kHz
QPOs; these typically occur in pairs)  are present in the HB and upper part
of the  NB.  Atoll sources exhibit  VLFN, HFN, 1--70~Hz QPOs  and kHz QPOs.
Atoll source HFN  can sometimes be peaked, and resemble  broad QPOs (HK and
references therein).  A recent review of the fast-time variability of Z and
atoll  sources, with  a discussion  of various  models, has  been  given by
van~der~Klis~(\cite{v00}).

In Z  sources, narrow QPOs (called  NBOs) with centroid  frequencies in the
range  5--8~Hz  have  been  observed  in the  NB.   NBOs  typically  become
detectable  halfway  down  the  NB,  and their  properties  differ  between
objects.  In  \object{Sco~X-1} the frequency of the  NBO is $\sim$5.5--8~Hz
in the NB, and jumps to $\sim$12~Hz as the source enters the FB (the QPO is
then referred to as  an FBO) (Dieters~\&~van~der~Klis~\cite{dv00}).  As the
source moves up the FB, the frequency and full width at half maximum (FWHM)
increase;  eventually, at  $\sim$10~$\%$  of the  way  up the  FB, the  FBO
becomes so broad as to  be indistinguishable from the underlying noise (but
the  relative  width never  exceeds  $\sim$0.5).   In \object{Sco~X-1}  the
fractional rms increases with increasing $\dot{M}$ and is greater at higher
energies;  and there  is no  time  lag between  low and  high energy  bands
(Dieters~et~al.~\cite{dvk00}).  The N/FBO  seen in \object{GX~17+2} behaves
in a similar fashion to \object{Sco~X-1} (Penninx~et~at.~\cite{plm90}).  In
the other  Z sources NBOs  have also been  observed but are  generally less
prominent (e.g. Kuulkers~et~al.~\cite{kvo94}).

\object{Sco~X-2}  is seemingly  very  similar to  \object{Sco~X-1}.  The  Z
tracks have  nearly identical shapes  (except for the  absence of an  HB in
\object{Sco  X-2})  and the  lightcurves  of  both  objects exhibit  strong
flaring               behaviour               (Schultz~et~al.~\cite{sht89};
Kuulkers~\&~van~der~Klis~\cite{kv95}).    The  orbital  periods   are  also
similar: 18.9~h  for \object{Sco~X-1} (Cowley~\&~Crampton~\cite{cc75}); and
$\sim$22~h for \object{Sco~X-2}  (Wachter~\cite{w97}).  One might therefore
expect the X-ray fast-time variability in \object{Sco~X-2} to resemble that
seen in \object{Sco~X-1}; but this is not the case.

EXOSAT      observations      of      \object{Sco~X-2}     reported      by
Ponman~et~al.~(\cite{pcs88})  showed  a complete  FB.   No  narrow NBO  was
found; however  broad peaked noise  was detected with a  centroid frequency
and  FWHM of  about  5~Hz and  10~Hz  respectively.  The  peaked noise  was
strongest  in the  lower FB,  with  a fractional  rms of  3.2$\pm$0.2~$\%$.
There was  no significant change  in the frequency  or width of  the peaked
noise, and there was no time lag between the low and high energy bands.  We
note that the authors divided their data into four segments on the basis of
intensity, and  not according to  position in the  Z track.  By  using that
method they could not differentiate between the NB and the lower FB.

A small  set of Rossi X-ray  Timing Explorer observations  were reported by
Kuulkers~\&~van~der~Klis~(\cite{kv98}; hereafter KK), when \object{Sco~X-2}
was near the NB/FB vertex.  In the  lower NB they found peaked noise with a
centroid frequency, FWHM and  fractional rms of 9.4$\pm$0.5~Hz, 16$\pm$2~Hz
and  2.7$\pm$0.1~$\%$ respectively.   In the  lower region  of the  FB they
measured    these   values    as   5.8$\pm$0.2~Hz,    11.0$\pm$0.7~Hz   and
4.2$\pm$0.1~$\%$.  There was a  small but significant decrease in frequency
when the  source moved through the  NB/FB vertex.  They  commented that the
peaked  noise resembled  the peaked  HFN sometimes  seen in  atoll sources.
They  found that the  peaked noise  fractional rms  was stronger  at higher
energies.

Even  though  there  have  been  clear  indications  that  \object{Sco~X-2}
exhibits marked  differences to  \object{Sco~X-1}, it has  remained largely
ignored.   This is perhaps  due to  the fact  that it  looks so  similar to
\object{Sco~X-1} in a CD and HID.

We  present results obtained  from an  analysis of  archived data  from the
Ginga satellite; these data have  not previously been presented.  This work
is  the  first  detailed  analysis  of the  power  spectral  properties  of
\object{Sco~X-2} as a  function of position in the  Z track.  This analysis
method  is   extremely  useful  because  it  allows   us  to  unambiguously
distinguish  between  the NB  and  lower FB,  and  is  helpful when  making
comparisons   between  different   objects.   We   further   highlight  the
differences  between \object{Sco~X-2}  and  \object{Sco~X-1}, and  conclude
that the observed peaked noise is  not the same as atoll source peaked HFN,
as  had been suggested.   We compare  our results  with the  power spectral
features seen in other Z sources.

\section{Observations \label{obs}}

The  Ginga  Large Area  Counter  (LAC) (Makino~\&~ASTRO-C  Team~\cite{m87};
Turner et al.~\cite{ttp89}) observed \object{Sco~X-2} between 5 and 7 March
1989.  We  retrieved  the data  from  the  Leicester  Database and  Archive
Service\footnote{see       http://ledas-www.star.le.ac.uk/ginga/}.       In
Table~\ref{tab:log} we present a log of observations.

The LAC  consisted of eight detectors,  and was sensitive to  X-rays in the
range  1--37~keV.  Several modes  were available,  with differing  time and
energy resolutions  (higher time resolution  sacrifices energy resolution).
The  observations discussed here  were done  in the  so-called MPC3  and PC
modes.  For  the MPC3 mode the  coarse gain setting was  high, providing an
energy  range 1--19~keV  over  twelve channels.   The  time resolution  was
7.8~ms  and  62.5~ms  for  high  (MPC3-H) and  medium  (MPC3-M)  bit  rates
respectively.  In the PC mode, the  LAC was divided into two groups of four
detectors each.   The coarse gain setting  was low, giving  an energy range
1.5--24.3~keV  for one  detector group,  and 1.2--17.9~keV  for  the other.
Each detector  group had two energy  channels, so the  PC mode observations
had four overlapping  energy bands.  In the high bit  rate mode (PC-H), the
time resolutions were  0.98~ms and 1.95~ms for the  lower and higher energy
bands respectively.  Our 0.98~ms data  were rebinned with a time resolution
of 1.95~ms, to obtain lightcurves for the full energy range.

During  the observations in  section C  (see Table~\ref{tab:log})  the high
voltages in  detectors 5 and 6  were accidently changed.   The energy range
for the  MPC3 observations stated  above is therefore only  approximate for
section C.

\begin{table*}[t]
\caption{Observation Log}
\label{tab:log}
\begin{tabular}{cccccc} 
\hline Section & Start Time (UT) &  End Time (UT) & Usable Data (s) & Modes
& Time Res (ms) \\ \hline

A & 5 Mar 1989 20:58 & 6 Mar 1989  00:17 & 4512 & PC-H & 0.98/1.95 \\ B & 6
Mar 1989 00:27  & 6 Mar 1989 19:13 & 2880  & MPC3-H & 7.8 \\ &  & & 12928 &
MPC3-M & 62.5 \\ C & 7 Mar 1989 02:13 & 7 Mar 1989 14:30 & 12928 & MPC3-M &
62.5 \\ D & 7  Mar 1989 18:12 & 7 Mar 1989 19:14 &  1184 & PC-H & 0.98/1.95
\\

\hline
\end{tabular}
\end{table*}

\section{Analysis \label{ana}}

The  counting rates  were corrected  for detector  deadtime  and collimator
response.  During the observations in sections A-C, the collimator response
was $\sim$0.96.  In  section D the response was  $\sim$0.73.  The data were
not  corrected  for  collimator   reflection.   The  effect  of  collimator
reflection was  to overestimate counting  rates below $\sim$6~keV by  a few
percent (Turner et al.~\cite{ttp89}; see also Hasinger~et~al.~\cite{hve90};
Wijnands~et~al.~\cite{wvk97}).

Hardness-intensity diagrams were constructed  using 64~s averages.  For the
MPC3  data we  defined  the hardness  as  the counting  rate ratio  between
8.2--16.7~keV and 5.8--8.2~keV.  The  intensity was defined as the counting
rate in the  range 1.1--16.7~keV.  For the PC data  we defined the hardness
as  the counting rate  ratio between  5.7--17.9~keV and  1.2--5.7~keV.  The
intensity was  defined as the counting  rate, from one  detector group only
(i.e. four detectors), in the range 1.2--17.9~keV.

Timing analysis was performed on  the raw (i.e.  not corrected for deadtime
and  collimator response) data.   Power spectra  were calculated  from 16~s
intervals in  several energy bands,  and then averaged according  to either
corrected counting rate or rank number.

Rank number was introduced by  Hasinger et al.~\cite{hve90} (see also Lewin
et        al.~\cite{llt92};        Hertz        et        al.~\cite{hvw92};
Dieters~\&~van~der~Klis~\cite{dv00})  and is  a one-dimensional  measure of
the position of the source in  a CD or HID.  Rank number increases smoothly
with increasing inferred mass-accretion rate.  To measure the rank numbers,
a spline was fitted through the  Z track, and each point was projected onto
it.   Rank number was  then determined  by measuring  the distance  of each
point along the spline.  We defined  the NB/FB vertex as rank number 1, and
the top of the observed FB as rank number 2.

We calculated  the expected  deadtime-affected white-noise levels  and then
subtracted the white-noise from each average power spectrum.  The procedure
to calculate white-noise levels in sum channel power spectra is well known,
and      is       described      by      van~der~Klis~(\cite{v89})      and
Mitsuda~\&~Dotani~(\cite{md89}).    The   full   procedure   to   calculate
white-noise levels in energy resolved power spectra is not explicity stated
in those references; we therefore describe our procedure in the Appendix.

After  the deadtime-affected  white-noise levels  had been  subtracted, the
spectra  were normalised  to fractional-rms-squared  per Hz  and functional
$\chi^{2}$  fits  were performed.   Very  low  frequency  noise (VLFN)  was
modelled using a  power law $A~\nu^{\alpha}$ and integrated  over the range
0.01--1  Hz.  Peaked  noise was  modelled  using a  Lorentzian.  All  power
spectra were initially  fitted with only a power  law.  The significance of
adding a Lorentzian component to the  fit was then estimated via an F-test.
If  the  inclusion  of  a   peaked  noise  component  was  not  significant
($<3\sigma$)  we   measured  90~$\%$  confidence  upper   limits  by  using
$\Delta\chi^{2}=2.71$.   Errors  on  all  parameters  were  measured  using
$\Delta\chi^{2}=1$.  When a  fit was not sensitive to  the centroid or FWHM
of the Lorentzian  (or was not significantly improved  by its inclusion) we
fixed those  parameters at  values measured from  other power  spectra (the
actual values are shown in Table~\ref{tab:mupper}).  We fitted each PC mode
power spectrum both with and without an extra free constant component.  The
reasons for doing this are discussed in Sect.~\ref{res_PC}.

All  fractional  rms  amplitudes  were  corrected  for  channel  cross-talk
(Lewin~et~al.~\cite{llt92}).  We assumed variations at all energies were in
phase.  Finally,  the fractional  rms values were  multiplied by  a binning
correction factor (van~der~Klis~\cite{v89}).

\section{Results \label{res}}

\subsection{Hardness-intensity diagrams \label{hids}} 

\begin{figure}[t]
\rotatebox{270}{\resizebox{!}{\hsize}{\includegraphics{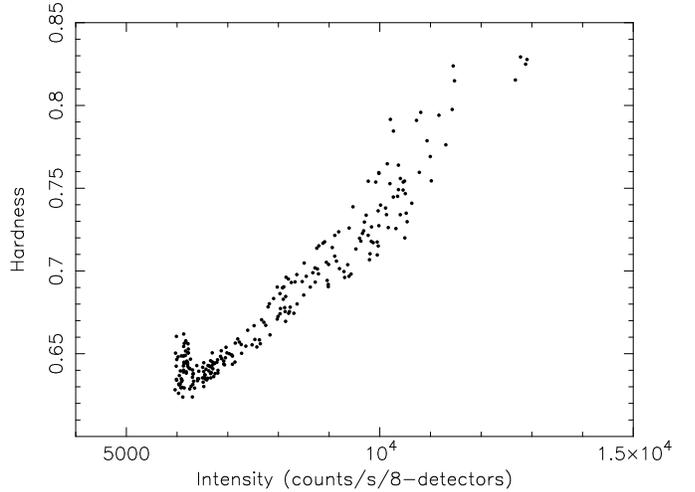}}}
\caption{HID of section B, with 64s averages. 
The  intensity  is the  counting  rate  in  the range  1.1--16.7~keV.   The
hardness is the counting rate ratio between 8.2--16.7~keV and 5.8--8.2~keV.
The error bars are slightly larger than the dots.}
\label{fig:hi_sectb}
\end{figure}

Fig.~\ref{fig:hi_sectb} shows  a HID  for all section  B data.   The small,
approximately vertical NB is clearly visible.  \object{Sco~X-2} is known to
flare to  intensities roughly twice  the level of the  persistent emission.
The highest intensity in Fig.~\ref{fig:hi_sectb} is a factor of 2.2 greater
than the lowest  intensity: we conclude therefore, that  we have observed a
fully developed flaring branch.  The  results from section C are compatible
with those  from section B; however  due to the uncertain  energy ranges of
those data, we will not discuss them any further.

\begin{figure}[t]
\rotatebox{270}{\resizebox{!}{\hsize}{\includegraphics{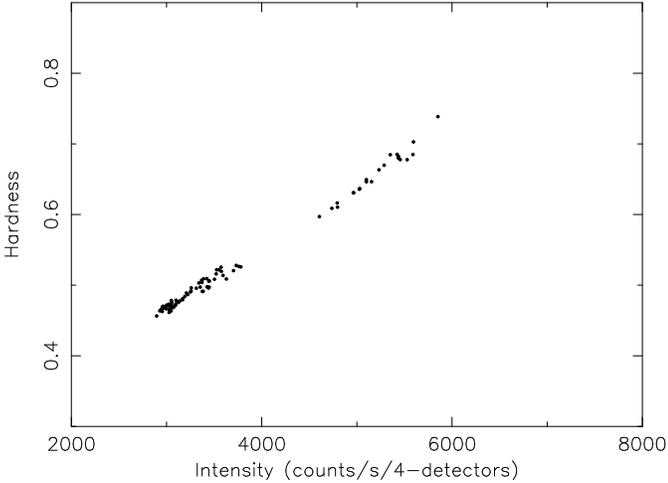}}}
\caption{HID of sections A and D, with 64~s averages. 
The  intensity  is the  counting  rate  in  the range  1.2--17.9~keV.   The
hardness   is   the  counting   rate   ratio   between  5.7--17.9~keV   and
1.2--5.7~keV. The error bars are slightly larger than the dots.}
\label{fig:fig_d_new}
\end{figure}

A HID of sections A and  D is shown in Fig.~\ref{fig:fig_d_new}. The lowest
and highest  intensities in Fig.~\ref{fig:fig_d_new} differ by  a factor of
2; therefore  (in comparison  with Fig.~\ref{fig:hi_sectb}), nearly  all of
the FB was traced out.  No clear NB/FB vertex is apparent (but we were able
to  estimate  where  it   might  lie;  see  Sect.~\ref{res_PC}).   When  we
constructed a HID  from MPC3 mode data using energy  bands similar to those
available in PC mode, the presence  of the NB was much less pronounced; the
NB became  more like a  `blob', rather than  a clear branch.   Therefore we
caution that in  Fig.~\ref{fig:fig_d_new} we cannot confidently distinguish
between data points from the lower NB (if there are any) and those actually
in the lower FB.  However, as  will be seen in Sect.~\ref{res_PC}, our main
interest  (with the  PC data)  is with  the  top half  of the  FB, so  this
ambiguity is not a major problem.

\subsection{Timing analysis of section B MPC3 mode data \label{res_H}} 

\begin{figure}[t]
\rotatebox{270}{\resizebox{!}{\hsize}{\includegraphics{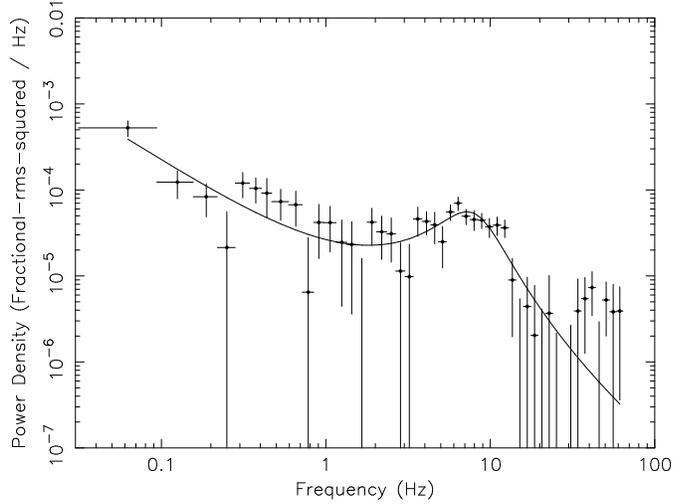}}}
\caption{MPC3-H power spectrum from the energy range 1.1--16.7~keV.  The
mean rank is 1.09 (lower FB).  The solid line shows the best fit.}
\label{fig:p063}
\end{figure}

During the  observations in MPC3-H mode,  the source was found  only in the
lower  FB.   We  divided  the  data  into  two  segments  with  mean  ranks
1.09$\pm$0.02  and   1.26$\pm$0.05.   An   example  of  a   power  spectrum
(1.1--16.7~keV)  is shown  in  Fig.~\ref{fig:p063}, with  mean rank  number
1.09.  The results of fitting the  MPC3-H data (1.1--16.7~keV) are shown in
Table~\ref{tab:mupper}. The FWHM of the best-fit Lorentzian is greater than
half its centroid frequency, so the feature cannot be formally described as
a QPO.   The MPC3-H data  show that there  is no significant change  in the
centroid frequency or FWHM of the  peaked noise, as the source moves up the
FB (the  decrease in the centroid  frequency between rank  numbers 1.09 and
1.26 is  significant only at  the $\sim$2~$\sigma$ level).   The fractional
rms  amplitude  of the  peaked  noise  is  also consistent  with  remaining
constant.

\begin{table*}[t]
\caption{Results from fits to the section B MPC3 power spectra from the
  energy range 1.1--16.7~keV.}
\label{tab:mupper}
\begin{threeparttable}
\begin{tabular}{ccccccc} \hline 
Mean Rank  & \multicolumn{2}{c}{VLFN} &  \multicolumn{3}{c}{Peaked Noise} &
Reduced $\chi^2$/dof \\ & Index & rms ($\%$) & Frequency (Hz) & FWHM (Hz) &
rms ($\%$) & \\ \hline

MPC3-H data \\

1.09$\pm$0.02 & -1.2$^{+0.2}_{-0.3}$  & 1.2$^{+0.2}_{-0.1}$ & 7.3$\pm$0.4 &
7.1$^{+1.2}_{-1.0}$ & 2.9$\pm$0.2 & 1.25 / 43 \\

1.26$\pm$0.05    &    -2.0$^{+0.2}_{-0.3}$    &    2.6$^{+0.6}_{-0.4}$    &
5.5$^{+0.8}_{-1.0}$ & 10.3$^{+3.1}_{-2.2}$ & 3.0$\pm$0.3 & 0.79 / 43 \\

MPC3-M data \\

0.95$\pm$0.02 & -1.5$\pm$0.3 &  1.2$\pm$0.2 & 4.3\tnote{a} & 6.1\tnote{a} &
1.8$\pm$0.2 & 0.85 / 25 \\

1.00$\pm$0.02 & -1.2$\pm$0.2 &  1.3$\pm$0.1 & 4.3\tnote{a} & 6.1\tnote{a} &
2.3$\pm$0.1 & 1.28 / 25 \\

1.09$\pm$0.03    &   -1.5$\pm$0.3   &    1.4$\pm$0.2   &    4.3$\pm$0.4   &
6.1$^{+2.1}_{-1.3}$ & 2.8$\pm$0.3 & 0.94 / 23 \\

1.27$\pm$0.06    &   -1.9$\pm$0.2   &    2.8$\pm$0.5   &    4.4$\pm$0.6   &
6.2$^{+2.9}_{-1.6}$ & 2.3$\pm$0.3 & 1.40 / 23 \\

1.41$\pm$0.03 & -1.6$\pm$0.1 &  2.9$\pm$0.2 & 4.4\tnote{a} & 6.2\tnote{a} &
1.4$\pm$0.2 & 0.84 / 25 \\

1.50$\pm$0.03 & -1.5$\pm$0.1 &  2.8$\pm$0.2 & 4.4\tnote{a} & 6.2\tnote{a} &
$<$1.3 & 1.06 / 26 \\

1.60$\pm$0.03 & -1.6$\pm$0.1 &  3.3$\pm$0.2 & 4.4\tnote{a} & 6.2\tnote{a} &
$<$0.9 & 1.10 / 26 \\

1.75$\pm$0.09 & -1.8$\pm$0.1 &  4.8$\pm$0.3 & 4.4\tnote{a} & 6.2\tnote{a} &
$<$1.2 & 1.10 / 26 \\ \hline

\end{tabular}
\begin{tablenotes}
\item[a] Fixed parameter
\end{tablenotes}
\end{threeparttable}
\end{table*}

We investigated the  energy dependence of the peaked  noise fractional rms,
by calculating energy resolved power spectra from the MPC3-H data.  We used
four energy bands: 1.1--3.4--5.8--8.2--16.7~keV.   All power spectra from a
particular band were  averaged together, giving a mean  power spectrum from
the lower  FB; the mean rank  number was 1.18$\pm$0.05.   We determined the
centroid frequency  and FWHM  of the peaked  noise feature from  an average
power spectrum  from the 1.1--16.7~keV  band: they were  6.7$\pm$0.4~Hz and
8.3$^{+1.2}_{-1.1}$~Hz respectively.  The centroid  and width were fixed at
those  values when fitting  the power  spectra from  the other  four energy
bands.  The  energy dependence of peaked  noise fractional rms  is shown in
Table~\ref{tab:h_lore}.   The peaked  noise is  clearly stronger  at higher
energies and is consistent with being constant above $\sim$6~keV.

\begin{table}[t]
\caption{Energy dependence of peaked noise fractional rms amplitude}
\label{tab:h_lore}
\begin{tabular}{cc} 
\hline Energy Range (keV) & rms ($\%$)\\ \hline

1.1--3.4 & $<$2.8 \\ 3.4--5.8 & $<$2.6 \\ 5.8--8.2 & 4.2$^{+0.3}_{-0.4}$ \\
8.2--16.7 & 4.8$^{+0.4}_{-0.5}$ \\

\hline
\end{tabular}
\end{table}

\begin{figure}[t]
\rotatebox{270}{\resizebox{!}{\hsize}{\includegraphics{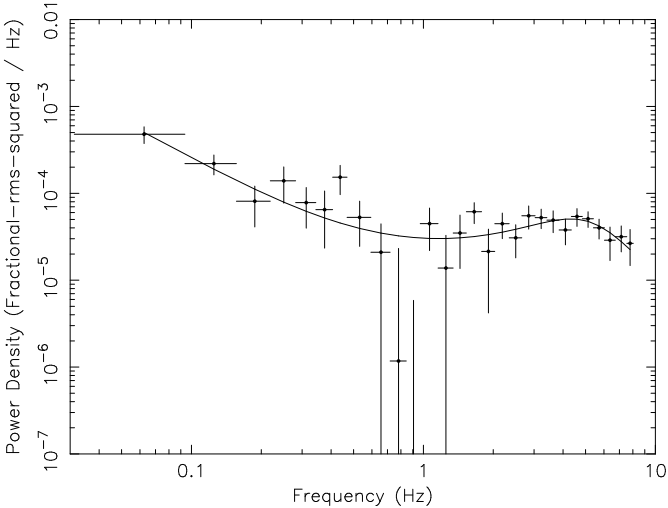}}}
\caption{MPC3-M power spectrum from the energy range 1.1--16.7~keV.  The
mean rank is 1.09 (lower FB).  The solid line shows the best fit.}
\label{fig:m_03}
\end{figure}

The  MPC3-M  data from  the  track  shown  in Fig.~\ref{fig:hi_sectb}  were
divided into eight  segments.  Power spectra were averaged  on the basis of
rank number.   An example (1.1--16.7~keV) is  shown in Fig.~\ref{fig:m_03},
with mean  rank number 1.09.   Table~\ref{tab:mupper} shows the  results of
fitting the  MPC3-M power spectra  (1.1--16.7~keV) from section B.   It was
only in the lower FB that  we could meaningfully constrain the centroid and
width  of   the  peaked  noise,   which  were  $\sim$4~Hz   and  $\sim$6~Hz
respectively.  There is no significant  change in those values between mean
ranks 1.09  and 1.27  (Table~\ref{tab:mupper}).  The centroid  frequency is
lower in  the MPC3-M  data than  in the MPC3-H  data.  This  discrepancy is
probably due to a lack of information above 8~Hz in the MPC3-M data.

\begin{figure}[t]
\rotatebox{270}{\resizebox{!}{\hsize}{\includegraphics{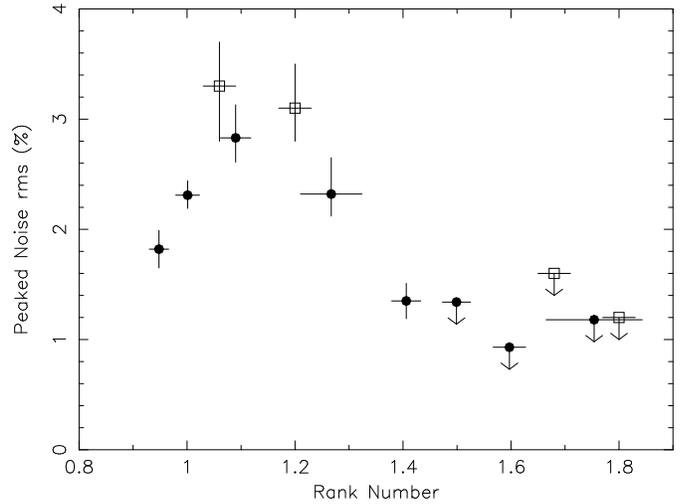}}}
\caption{Peaked noise fractional rms values as a function of rank number
for the MPC3-M data (solid dots, 1.1--16.7~keV), and PC data (open squares,
1.2--17.9~keV).  The  fractional rms  values of the  PC data are  from fits
which included a constant component.}
\label{fig:med_lore_B}
\end{figure}

\begin{figure}[t]
\rotatebox{270}{\resizebox{!}{\hsize}{\includegraphics{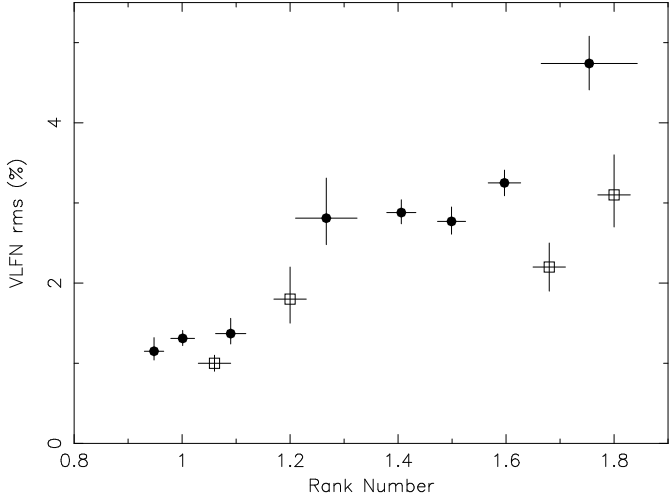}}}
\caption{VLFN fractional rms values as a function of rank number for the
MPC3-M  data  (solid  dots,  1.1--16.7~keV),  and PC  data  (open  squares,
1.2--17.9~keV).  The  fractional rms  values of the  PC data are  from fits
which included a constant component.}
\label{fig:med_vlfn_B}
\end{figure}

The dependence  of peaked noise and  VLFN fractional rms on  rank number is
shown  in Figs.~\ref{fig:med_lore_B}~and~\ref{fig:med_vlfn_B} respectively.
The peaked  noise fractional rms  is greatest in  the lower FB  and becomes
undetectable (upper limit 1.3~$\%$) by about halfway up the FB.  Of note is
the fact that  it \emph{is} actually detectable up to a  mean rank of 1.41.
(This  is   important  because  N/FBO  typically   become  undetectable  by
$\sim$10~$\%$ of the way up the FB.)  The drop in power as the source moves
further up the FB or into the NB may be due to the feature moving above the
Nyquist frequency  of these data; however  we can exclude  this effect with
respect  to  movement  up   the  FB  (see  Sect.~\ref{res_PC}).   The  VLFN
fractional  rms is  seen to  increase as  the source  moves up  the  FB, as
expected  for Z sources  (HK). The  slope of  the power  law appears  to be
uncorrelated with rank number.

\subsection{Timing analysis of PC mode data \label{res_PC}}

The NB/FB  vertex was not unambiguously  observed, so we  did not calculate
rank numbers for each individual  interval; averaging was done on the basis
of  intensity.  The  track was  divided into  four segments;  the intensity
ranges  were:  2800--3200,  3200--3900, 4500--5200  and  5200--5900~count/s
(four detectors only).   The mean intensities of those  four segments were:
3041$\pm$6, 3458$\pm$14,  4962$\pm$25 and 5492$\pm$31~counts/s.   We could,
however, estimate  rank numbers by  comparing Fig.~\ref{fig:fig_d_new} with
Fig.~\ref{fig:hi_sectb} (assuming the intensities are derived from the same
energy  band).    As  stated  in  Sect.~\ref{hids},  the   track  shown  in
Fig.~\ref{fig:fig_d_new} is not quite  a complete FB.  The lowest intensity
observed  in Fig.~\ref{fig:fig_d_new} is  $\sim$2900~counts/s; this  can be
taken  as  the   upper  limit  of  the  intensity   of  the  NB/FB  vertex.
Alternatively,   if   we   assume   the  highest   intensity   present   in
Fig.~\ref{fig:fig_d_new} is  actually the  top of the  FB, then  the unseen
NB/FB vertex can  be placed at $\sim$2780~counts/s; this  can be considered
to be the lower limit.  By considering these two extremes we calculated the
approximate rank  numbers of  the four segments  as: 1.06, 1.20,  1.68, and
1.80  ($\pm$0.03).  We  stress that  these rank  numbers are  only  a rough
guide, and that the lower FB in Fig.~\ref{fig:fig_d_new} may infact contain
data from the NB.

\begin{figure}[t]
\begin{center}
\rotatebox{270}{\resizebox{!}{\hsize}{\includegraphics{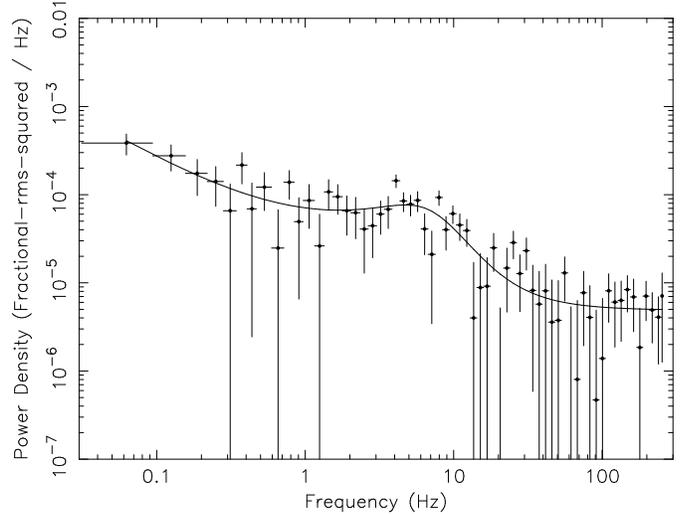}}}
\caption{PC power spectrum from the energy range 1.2--17.9~keV.  The
mean intensity  is 3041~counts/s (mean  rank number $\sim$1.06;  lower FB).
The solid line shows the best fit, and includes a constant component.}
\label{fig:n_20_01_C}
\end{center}
\end{figure}

An   example   power   spectrum   is  shown   in   Fig.~\ref{fig:n_20_01_C}
(1.2--17.9~keV),  with  mean  intensity  3041~counts/s  (mean  rank  number
$\sim$1.06).  The results from fitting the PC power spectra (1.2--17.9~keV)
are shown in Table~\ref{tab:pcfits}, where  we present the fits from models
both with  and without a constant.   The results from fits  with a constant
are shown also in Figs.~\ref{fig:med_lore_B} and \ref{fig:med_vlfn_B}.  The
most important  thing to note  is the nondetection  of peaked noise  in the
upper FB (upper~limit~1.6~$\%$).  The  PC data Nyquist frequency is 256~Hz;
therefore the nondetection of peaked noise in these data is not due only to
the feature moving to higher  frequencies.  From this, we conclude that the
nondetection of peaked noise in the MPC3-M data (Table~\ref{tab:mupper}) is
also a real weakening of the peak.

\begin{table*}[t]
\caption{Results from fits to the PC power spectra from the energy range
  1.2--17.9~keV.}  {\scriptsize
\label{tab:pcfits}
\begin{threeparttable}
\begin{tabular}{cccccccc} \hline 
Mean Intensity& \multicolumn{2}{c}{VLFN} & \multicolumn{3}{c}{Peaked Noise}
& Constant  & Reduced  $\chi^2$/dof \\ (cnts/s/4-detectors)  & Index  & rms
($\%$) & Frequency (Hz) & FWHM (Hz) & rms ($\%$) \\ \hline

Fits without an extra constant \\

3041$\pm$6   &   -0.5$\pm$0.1  &   1.1$\pm$0.1   &  5.4$^{+0.9}_{-0.7}$   &
7.2$^{+2.7}_{-3.1}$ & 2.5$\pm$0.4 & --- & 1.11 / 58 \\

3458$\pm$14 &  -0.5$\pm$0.1 & 1.2$\pm$0.1  & 5.4\tnote{a} &  7.2\tnote{a} &
$<$3.1 & --- & 1.26 / 61 \\

4962$\pm$25 &  -1.3$\pm$0.2 & 2.0$\pm$0.3  & 5.4\tnote{a} &  7.2\tnote{a} &
$<$1.8 & --- & 1.23 / 61 \\

5492$\pm$31   &  -1.4$\pm$0.2  &   2.7$^{+0.5}_{-0.4}$  &   5.4\tnote{a}  &
7.2\tnote{a} & $<$1.3 & --- & 1.44 / 61 \\

Fits with an extra constant \\

3041$\pm$6  & -1.0$^{+0.2}_{-0.4}$  & 1.0$\pm$0.1  &  5.0$^{+0.8}_{-1.0}$ &
10.3$^{+3.0}_{-2.5}$            &           3.3$^{+0.4}_{-0.5}$           &
4.8e-6$^{+1.0\mathrm{e}-6}_{-1.3\mathrm{e}-6}$ & 1.04 / 57 \\

3458$\pm$14  & -1.7$\pm$0.3 &  1.8$^{+0.4}_{-0.3}$ &  5.9$^{+1.0}_{-1.1}$ &
11.9$^{+4.1}_{-3.2}$ & 3.1$^{+0.4}_{-0.3}$ &  5.3e-6$\pm$1.0e-6 & 0.58 / 57
\\

4962$\pm$25 &  -1.5$\pm$0.2 & 2.2$\pm$0.3 & 5.9\tnote{a}  & 11.9\tnote{a} &
$<$1.6 & 3.6e-6$\pm$1.1e-6 & 1.06 / 60 \\

5492$\pm$31   &  -1.6$\pm$0.2  &   3.1$^{+0.5}_{-0.4}$  &   5.9\tnote{a}  &
11.9\tnote{a} & $<$1.2 & 4.3e-6$\pm$0.9e-6 & 1.10 / 60 \\ \hline
\end{tabular} 
\begin{tablenotes}
\item[a] Fixed parameter
\end{tablenotes}
\end{threeparttable}
}
\end{table*}

The   reduced    $\chi^2$s   of   the   fits   without    a   constant   in
Table~\ref{tab:pcfits} are  formally acceptable; however  visual inspection
of the  power spectra revealed excess  power above about  100~Hz (e.g.  see
Fig.~\ref{fig:n_20_01_C}), and  the best-fit VLFN indices found  in the two
lowest intensity segments are  unusually flat.  Given these indications, we
attempted to  account for  a possible high  frequency feature.  We  added a
cut-off power law  to our model and attempted to fit  the spectrum from the
lowest  intensity segment.   We could  not constrain  the index  or cut-off
frequency of this  component.  When we fitted a  Lorentzian (instead of the
cut-off  power law) to  the same  spectrum we  could constrain  neither the
width nor centroid frequency.  The  fits were improved with the addition of
the  extra component,  but were  not sensitive  to the  exact form  of that
component.  We  fitted the spectrum with  a constant (instead  of a cut-off
power law or  Lorentzian) and the fit also improved.   None of the best-fit
values of the peaked noise or VLFN were sensitive to the type of model that
we used to describe the excess power; for simplicity, we therefore chose to
use a constant.

In  the three  highest  intensity  segments, the  inclusion  of a  constant
significantly improved the fits (determined via an F-test for the inclusion
of the extra  component).  Furthermore, in the segment  with mean intensity
3458~counts/s, fitting  without a constant resulted only  in determining an
upper limit  on the  peaked noise of  $<$3.1~$\%$ (Table~\ref{tab:pcfits}).
This upper  limit is  rather large,  and a visual  inspection of  the power
spectrum showed peaked noise to  be present.  Therefore, it seems justified
to include an extra component in the fits, without needing (nor being able)
to describe the  actual form of the possible  higher frequency feature.  We
note that:  HFN was reported in \object{Sco~X-2},  with cut-off frequencies
of 66~Hz and  41~Hz in the NB and  FB respectively (HK); also a  QPO with a
frequency  of  $\sim$125~Hz  and   FWHM  of  $\sim$30~Hz  was  detected  by
Penninx~et~al.~(\cite{plm90}), in the FB of \object{GX~17+2}.

The fact that we cannot determine  the form of the higher frequency feature
is not a  major problem.  The main  result we obtained from the  PC data is
this: the  peaked noise feature experienced  a real weakening  in the upper
FB.  This result is not sensitive  to the presence or otherwise of a higher
frequency feature.

\section{Discussion \label{dis}}

\subsection{Observed properties of \object{Sco~X-2} \label{dis_prop}}

We were able to trace the  power spectral features of \object{Sco~X-2} as a
function of position  in the Z track, in  the energy range $\sim$1--17~keV.
At  no time  did we  detect a  narrow N/FBO  as found  in \object{Sco~X-1}.
Broad peaked noise was  present at $\sim$4--7~Hz, with FWHM $\sim$6--12~Hz.
We will subsequently  refer to this feature as  flaring branch noise (FBN).
FBN was detected in the lower NB  and reached a maximum at rank number 1.09
($\sim$10~$\%$ of the  way up the FB).  At its  maximum, the fractional rms
was  $\sim$3~$\%$.  It  then became  weaker  and was  undetectable by  rank
number  1.5 (the  middle of  the FB;  upper limit  1.3~$\%$).  We  found no
significant change (more  than a few~Hz) in the  centroid frequency or FWHM
as the source moved up the FB.  The FBN was stronger at higher energies.

The  behaviour  we  have observed  is  very  similar  to that  reported  by
Ponman~et~at.~(\cite{pcs88}).    In  addition,   we  have   been   able  to
unambiguously separate  NB data  from lower FB  data.  The increase  in the
fractional rms that  accompanied movement from the NB  through to the lower
FB, was also seen in the RXTE data of KK.  We have not been able to confirm
the decrease that  they observed in the centroid  frequency and FWHM, which
accompanied that transition.

\subsection{Comparison with other power spectral features \label{dis_comp}}

The fast-time variability of \object{Sco~X-2}  is different to that seen in
the  other Z  sources.   Most obvious  is  the low  coherence  of FBN  when
compared with N/FBOs.  The FBN has a relative width of $\sim$1--2, compared
with the N/FBO in \object{Sco~X-1} which  has a relative width of less than
$\sim$0.5 (Dieters~\&~van~der~Klis~\cite{dv00}).  Also in \object{Sco~X-1},
the frequency  and FWHM of the  NBO increase with  increasing inferred mass
accretion  rate, and  there is  a  sudden jump  in frequency  at the  NB/FB
vertex.  In \object{Sco~X-2} on the other  hand, there is a decrease in the
FBN centroid frequency and FWHM  with movement through the vertex (KK).  In
\object{Sco~X-1} the  NBO strength  increases with increasing  rank number;
the   NBO  becomes   indistinguishable  from   the  underlying   noise  (by
$\sim$10~$\%$  of  the way  up  the FB)  because  the  FWHM increases.   In
comparison, FBN  is strongest at  about 10~$\%$ of  the way up the  FB, and
becomes  undetectable because of  a subsequent  decrease in  the fractional
rms.

Even given the marked differences,  there are some similarities between FBN
and  the  N/FBO seen  in  \object{Sco~X-1}:  the  centroid frequencies  are
similar, both the N/FBO and FBN are  hard, and there is no time lag between
the soft and hard energies  in either phenomenon.  In \object{Sco~X-1}, the
NB/FB vertex  is a critical position  in the Z  track.  In \object{Sco~X-2}
there are hints  that this may also be true, although  in a different sense
(i.e.  frequency \emph{jump} for NBOs compared with a frequency \emph{fall}
for  FBN).  FBN  and  Z source  N/FBOs  may still  be  related, as  already
suggested by KK.

In the scheme of the unified model, radiation pressure drives matter up out
of the  disc, forming an  inner disc corona,  resulting in a  radial inflow
(Lamb~\cite{l91};  Psaltis~et~al.~\cite{plm95}).  When  $\dot{M}$  rises to
within  $\sim$20~$\%$ of the  Eddington limit,  radial oscillations  in the
optical  depth of the  inflowing material  produce a  rocking of  the X-ray
spectrum;   this   gives   rise   to   NBOs   (Fortner~et~al.~\cite{flm89};
Miller~\&~Park~\cite{mp95}).  This  model may explain the  evolution of the
N/FBO seen in \object{Sco~X-1}, but it  does not do well in explaining FBN.
The FBN  centroid frequency  does not increase  as predicted by  the model.
Furthermore,  we have  detected  it $\sim$40~$\%$  of  the way  up the  FB;
oscillations are predicted to be suppressed at such high accretion rates.

QPOs  at  super-Eddington  accretion   rates  have  been  observed  before:
Kuulkers~\&~van~der~Klis~(\cite{kv95b}) observed  a $\sim$26~Hz QPO  in the
upper FB of \object{Cyg~X-2}.  The QPO was observed during an intensity dip
in the FB,  and it was stronger at lower energies.   They suggested that it
is produced  by oscillations in an  inner disc torus, and  it is observable
because   \object{Cyg~X-2}   is  viewed   at   a   high  inclination   (see
Sect.~\ref{dis_type}).  \object{Sco~X-2}, on the  other hand, is thought to
be viewed at a low  inclination (see Sect.~\ref{dis_type}); therefore it is
unlikely that FBN is related to the 26~Hz QPO.

KK  pointed out  that the  FBN seen  in \object{Sco~X-2}  looks  similar to
peaked HFN sometimes seen in  atoll sources.  We observed a \emph{decrease}
in the FBN fractional rms as the  source moved up the NB, which is opposite
to what would be expected if the  FBN was infact atoll source HFN.  The low
Nyquist  frequency  is  no problem  here  as  the  break frequency  of  HFN
decreases     with      decreasing     inferred     mass-accretion     rate
(Wijnands~\&~van~der~Klis~\cite{wv99}).  The  phenomenology of FBN  is thus
significantly different to atoll source  HFN; therefore we propose that FBN
is unrelated to atoll source HFN.

\subsection{Two types of Z sources: a clue to the origin of FBN?\label{dis_type}}

In order to  understand why \object{Sco~X-2} behaves so  differently, it is
worthwhile to investigate the  general relationships that exist between the
Z  sources.   On  the basis  of  their  Z  track morphology  and  fast-time
variability,  they  can  be  divided  into  two  groups:  \object{Cyg~X-2},
\object{GX~5-1},     \object{GX~340+0}      (Cyg-like     objects);     and
\object{Sco~X-1},  \object{Sco~X-2},  \object{GX~17+2}  (Sco-like  objects)
(e.g. Kuulkers~et~al.~\cite{kvo97}).  The Cyg-like sources have much larger
HBs and  smaller FBs than  the Sco-like sources.   The HBs in  the Sco-like
objects are slanted, while the  HBs in the Cyg-like sources are horizontal.
Most importantly, in the Cyg-like objects  the intensity is seen to fall in
the middle and upper FB, while  the Sco-like objects really do flare in the
FB.

At high accretion rates an inner  disc torus may be produced, which in high
inclination  sources can  partly obscure  the X-ray  emitting  region.  The
Cyg-like  objects have  thus been  interpreted  as being  viewed at  higher
inclinations,   and   the    Sco-like   objects   at   lower   inclinations
(Kuulkers~et~al.~\cite{kvo97}).  In  support of this,  optical observations
place   \object{Cyg~X-2}    at   an   inclination    of   62.5$\pm$4$\degr$
(Orosz~\&~Kuulkers~\cite{ok99}),  \object{Sco~X-1}  at 15$\degr$--40$\degr$
(Crampton~et~al.~\cite{cch76}), and  also suggest that  \object{Sco~X-2} is
viewed    at    a    low    inclination    (Wachter~\&~Margon~\cite{wm96}).
Psaltis~et~al.~(\cite{plm95}) have modelled the X-ray spectra of Z sources.
They suggested that  the Sco-like objects have weaker  magnetic fields than
the Cyg-like objects.

The  most  obvious difference  between  \object{Sco~X-2}  and  the other  Z
sources is the  absence of an HB,  so it is tempting to  think that either:
\object{Sco~X-2} has  a weaker magnetic field  than \object{Sco~X-1}, while
also (perhaps  incidentally) being viewed  at a low inclination;  or simply
that  \object{Sco~X-2} never reaches  a low  enough mass-accretion  rate to
trace out an HB.

A  close investigation  of the  timing properties  of  \object{Cyg~X-2} may
provide  hints as  to the  origin  of FBN.   In \object{Cyg~X-2}  long-term
variations  in  the  mean   X-ray  intensity  have  been  discovered  (e.g.
Paul~et~al.~\cite{pkm00}).   Ginga observations of  \object{Cyg~X-2} showed
that the  Z track was more  Sco-like at medium level  intensities, and more
Cyg-like  at high  level intensities  (Wijnands~et~al.~\cite{wvk97}).  RXTE
observations of \object{Cyg~X-2} at  low intensities revealed no NBO (upper
limit $\sim$1~$\%$) in the NB;  however a peaked noise feature was detected
in  the lower  FB,  with a  frequency  and fractional  rms  of 6--7~Hz  and
$\sim$3~$\%$ respectively (Kuulkers~et~al.~\cite{kwv99}).  This is (perhaps
only superficially) similar to the FBN seen in \object{Sco~X-2}.

It  is  vital  to  have   more  observations  of  \object{Cyg~X-2}  at  low
intensities so that  we can better understand the  properties of the peaked
noise  feature.   Only   then  can  we  make  any   conclusions  about  the
relationship between FBN and the \object{Cyg~X-2} peaked noise.  But if the
two features  are the same phenomenon,  we must ask: what  are the physical
characteristics of  \object{Cyg~X-2} when it  is at low  level intensities?
The low level may be due to obscuration by the outer
accretion    disc,     possibly    caused    by     a    precessing    disc
(e.g.                                        Kuulkers~et~al.~{\cite{kvv96}).
Wijnands~\&~van~der~Klis~(\cite{wv01})  have argued against  the precessing
disc  model on  the  basis of  differing  kHz QPO  properties at  different
intensity levels.  Kuulkers~et~al.~(\cite{kwv99}) showed that the relative
VLFN properties at different intensity levels were also inconsistent with
obscuration.   Instead,  they  preferred  a  model  in  which  the  overall
intensity level is sensitive to the properties of the inner mass flow.
Whatever the case, if FBN and \object{Cyg~X-2} peaked noise are produced by
the same process, then that  process will be largely insensitive to orbital
inclination.  In this  way, \object{Sco~X-2} may prove to  be a useful tool
in  discerning   between  models  trying   to  explain  the   behaviour  of
\object{Cyg~X-2}, and ultimately the other Z sources.

\section{Conclusions \label{con}}

We  have  carried out  the  first detailed  study  of  the X-ray  fast-time
variability of the Z source  \object{Sco~X-2}, as a function of position in
the Z track.  We found broad  peaked noise with centroid frequency and FWHM
in  the range $\sim$4--7~Hz  and $\sim$6--12~Hz  respectively.  We  find it
difficult to attribute FBN to the  same processes which are able to explain
typical Z source  N/FBOs.  We also conclude that it  is not a manifestation
of atoll source HFN.   FBN resembles the FB peaked noise seen  in the FB of
\object{Cyg~X-2} at low overall  intensities.  If these are later confirmed
to be  the same  phenomena, then  the process responsible  for FBN  must be
largely insensitive to orbital inclination.

\FloatBarrier

\begin{acknowledgements}

This research has made use of data obtained from the Leicester Database and
Archive  Service at  the  Department of  Physics  and Astronomy,  Leicester
University, UK.   The reduction  of raw data  was carried out  with support
from  the Cooperative  Research Centre  for Advanced  Computational Systems
established under the  Australian Government's Cooperative Research Centres
Program.  PMO  wishes to thank the Space  Research Organization Netherlands
for its  hospitality, and  is currently being  supported by  the Australian
Postgraduate Award scheme.

\end{acknowledgements}

\appendix
\section*{Appendix: Calculation of deadtime-affected white-noise levels}
\Alph{section} \setcounter{section}{1}

In the  absence of  deadtime, the white-noise  level in a  Leahy normalised
 power  spectrum,  due  to  Poisson  noise  in the  data,  is  equal  to  2
 (Leahy~et~al.~\cite{lde83}).  The  effect of  deadtime is to  decrease the
 variance  of  the  data  and   thus  lower  the  white-noise  level.   The
 deadtime-affected   white-noise  level   for  the   sum   channel  $P_{\nu
 \mathrm{s}}$,  can be  calculated for  each frequency  $\nu$ in  the power
 spectrum

\begin{multline}
<P_{\nu                            \mathrm{s}}>                           =
2\left(1-\mu_{\mathrm{s}}\tau_{\mathrm{dead}}\right)^{2}\times            \\
\left(1+2\left(\frac{\mu_{\mathrm{s}}\tau_{\mathrm{dead}}}{1-\mu_{\mathrm{s}}\tau_{\mathrm{dead}}}\right)\left(\frac{\tau_{\mathrm{dead}}}{\tau_{\mathrm{res}}}\right)\sin^{2}\left(\frac{\pi\nu}{2\nu_{\mathrm{Nyquist}}}\right)\right)
\end{multline}

\noindent where $\mu_{\mathrm{s}}$ is the observed counting rate per detector,
$\tau_{\mathrm{res}}$   is   the  time   resolution   of  the   lightcurve,
$\tau_{\mathrm{dead}}$     is    the     deadtime     per    event,     and
$\nu_{\mathrm{Nyquist}}$  is  the   Nyquist  frequency  of  each  transform
(Weisskopf~\cite{w85}).

The  deadtime-affected white-noise  level  for any  particular energy  band
$P_{\nu   \mathrm{b}}$,    can   be    calculated   as   a    function   of
$P_{\nu\mathrm{s}}$.   For an  arbitrary time  interval, the  ratio  of the
total number  of counts  $k$ in  the particular energy  band, to  the total
number  of counts  $n$ in  the sum  channel is  denoted by  $r$.   The mean
variance  of   the  data   is  denoted  as   $\sigma^{2}_{\mathrm{s}}$  and
$\sigma^{2}_{\mathrm{b}}$  for the  sum channel  and the  particular energy
band respectively.  These are related by (Mitsuda \& Dotani~\cite{md89})

\begin{equation}
\sigma^{2}_{\mathrm{b}} = r^{2}\sigma^{2}_{\mathrm{s}} + k\left(1-r\right)
\label{eq:1}
\end{equation}

Each transform is  calculated from a time series  $x_{k}$ ($k = 0,...,N-1$)
consisting  of  $N$  arbitrary  time intervals.   By  employing  Parseval's
theorem, we  can relate the total variance  $Var\left(x_{k}\right)$ (or the
mean variance $\sigma^{2}$)  of the time series, and  the Fourier amplitude
$a$

\begin{equation}
Var\left(x_{k}\right) = N\sigma^{2} = \frac{N-1}{N}|a|^{2}
\end{equation}

\noindent Here, we have considered only situations where the dependence of
Fourier  amplitude  on frequency  is  negligible  (i.e. white-noise).   The
factor of N-1 originates from the  fact that the sum used in the expression
for the total variance does not include a term for a frequency of zero.

The mean variance $\sigma^2$ and Leahy power $P$ can then be related via

\begin{equation}
\sigma^{2} = P\frac{N_{\mathrm{ph}}}{2}\left(\frac{N-1}{N^{2}}\right)
\label{eq:2}
\end{equation}

$N_{\mathrm{ph}}$ is  equal to  $Nn$ for  the sum channel  or $Nk$  for the
particular energy band.   We can express $P_{\mathrm{b}}$ as  a function of
$P_{\mathrm{s}}$ by substituting Eq.~\ref{eq:2} into Eq.~\ref{eq:1}
  
\begin{equation}
P_{\mathrm{b}}\frac{k}{2}\left(\frac{N-1}{N}\right)=
      P_{\mathrm{s}}\frac{k^{2}}{n^{2}}\frac{n}{2}\left(\frac{N-1}{N}\right)
      + k\left(1-r\right)
\end{equation}

\noindent or

\begin{equation}
P_{\mathrm{b}}           =           2\left(\frac{N}{N-1}\right)          -
\frac{k}{n}\left(2\left(\frac{N}{N-1}\right)-P_{\mathrm{s}}\right)
\end{equation}

\noindent which in the case of large $N$ reduces to

\begin{equation}
P_{\mathrm{b}} = 2 - \frac{k}{n}\left(2 - P_{\mathrm{s}}\right)
\end{equation}

\end{document}